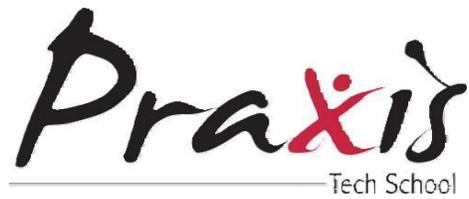

# Boosting Digital Safeguards: Blending Cryptography and Steganography

A Capstone Project report submitted in partial fulfillment of the requirements for the Post Graduate Program in Data Science at Praxis Tech School, Kolkata, INDIA

By

Anamitra Maiti (A23003)
Subham Laha (A23045 )
Rishav Upadhaya(A23033)
Soumyajit Biswas(A23044)
Vikas Chaudhary(A23048)
Biplab Kar(A23014)
Nikhil Kumar(A23026)

Under the supervision of
Prof. Jaydip Sen
Praxis Tech School, Kolkata, India

# Boosting Digital Safeguards: Blending Cryptography and Steganography

Anamitra Maiti[1], Subham Laha[2], Soumyajit Biswas[3], Rishav Upadhaya[4], Vikas Chaudhary[5], Nikhil Kumar[6], Biplab Kar[7], Jaydip Sen[8]
Email: {[1]anamitra.maiti_ds23fall, [2]subham.laha_ds23fall, [3]soumyajit.biswas_ds23fall, [4]rishav.upadhaya_ds23fall, [5]vikas.chaudhary_ds23fall, [6]nikhil.kumar _ds23fall, [7]biplab.kar_ds23fall}@praxistech.school, [8]jaydip@praxis.ac.in

## 1. Introduction

Prominent methods for modifying data to either encrypt or hide its provenance include cryptography and steganography, respectively. Steganography involves the art and science of covert communication, effectively making the communication invisible. Conversely, cryptography encrypts a message, rendering it unintelligible. While each method offers its own form of security, research has explored integrating both to enhance confidentiality and overall security. Steganography involves the art and science of covert communication, effectively making the communication invisible. Conversely, cryptography encrypts a message, rendering it unintelligible. While each method offers its own form of security, research has explored integrating both to enhance confidentiality and overall security.

Cryptographic systems are generally categorized into symmetric-key systems, which use an individual shared key between the sending and receiving parties, and public-key systems, which use two keys: a private key that is utilized by the recipient of the message and a public key that is known to everyone. In the context of cryptography, a ciphered message might raise suspicion, whereas a message hidden through steganographic techniques remains undetected. The effectiveness of steganography versus cryptography is measured differently: steganography is compromised if the hidden content is accessed, whereas cryptography fails if the presence of a secret message is detected. The fields dedicated to breaking encrypted information and uncovering hidden messages are known as cryptanalysis and steganalysis, respectively, focusing on decrypting information and detecting concealed messages .The effectiveness of steganography versus cryptography is measured differently: steganography is compromised if the hidden content is accessed, whereas cryptography fails if the presence of a secret message is detected. The fields dedicated to breaking encrypted information and uncovering hidden messages are known as cryptanalysis and steganalysis, respectively, focusing on decrypting information and detecting concealed messages.

This paper aims to outline an approach for merging cryptography and steganography using images. As essential components of computer security, concealment, access, and validity are shared goals that both cryptography and steganography strive to achieve. These methods facilitate the secure transmission of private data over public networks, allowing only those with the secret key to access the encrypted messages, which could range from documents to images or other data types. These methods facilitate the secure transmission of private data over public networks, allowing only those with the secret key to access the encrypted messages, which could range from documents to images or other data types.

Cryptography and steganography not only play vital roles in computer and network security, particularly in access control and protecting information confidentiality, but they are also integral to many daily applications. Despite their differences, the demand for both cryptography and steganography has surged with the rapid expansion of the Internet, highlighting their importance in the digital age.

## 2. Related Work

Our project is at the intersection of three cutting-edge fields: cryptographic algorithms, steganography, and Generative Adversarial Networks (GANs). The goal is to enhance digital security by leveraging the unique capabilities of these technologies. In this section, we review seminal literature and prior research that forms the basis of our project's core and direction.

Recent studies have delved into the use of GANs to embed hidden messages within images. GANs, comprising two neural networks in competition, are employed to generate images that appear realistic while containing concealed information. This approach represents a novel method in steganography, where GANs play a pivotal role in concealing messages within digital images.

Further advancements in steganography have been proposed through the development of specialized GANs. These networks are trained to encode secret messages directly into images, enhancing the traditional methods of steganography. This innovative approach expands the capabilities of concealing information within digital media.

Literature reviews have extensively explored techniques for hiding messages within digital images. These techniques range from subtle modifications of pixel values to complex mathematical transformations, with a notable emphasis on the increasing utilization of GANs in image steganography. Such comprehensive surveys provide invaluable insights into the evolving landscape of steganographic methods.

Moreover, research has conducted in-depth analyses of digital image steganographic techniques, categorizing them based on their approaches to modifying images and embedding messages. Additionally, discussions on steganalysis, the detection of hidden messages within images, underscore the importance of understanding both concealment and detection strategies in digital security.

Beyond steganography, studies have surveyed the broader applications of GANs in computer security. These applications include not only steganography but also privacy protection through the generation of realistic synthetic data and the creation of adversarial examples to assess the robustness of machine learning models.

Innovations in encryption techniques have also been explored, particularly in using generative networks to encrypt images while preserving visual quality. This addresses the challenge of traditional encryption methods, which may distort the appearance of encrypted images.

Furthermore, research has investigated the use of deep learning, specifically attention mechanisms, to detect hidden messages within images generated using GANs. This represents a dual use of GAN technology, both for concealing and revealing hidden information.

Additionally, advancements in steganographic methods have been proposed, such as an AI-enhanced interface for Least Significant Bit (LSB) steganography. This interface aims to bolster the resilience and safety associated with LSB steganography, which is vulnerable to detection due to its simplicity.

Other approaches combine LSB steganography with encryption to enhance the security of hidden messages within images. These methods address the limitations of LSB steganography by adding an extra layer of encryption for heightened security.

In the realm of cryptographic algorithms, research has proposed more efficient key exchange protocols and hybrid cryptosystems combining aspects of different encryption techniques. These innovations aim to enhance security and efficiency in cryptographic operations.

Overall, the insights gleaned from the reviewed literature provide a robust theoretical and practical foundation for our project. By integrating principles from cryptography, steganography, and GAN technology, we aim to develop a sophisticated security system characterized by unparalleled security, stealth, and efficiency. These studies not only inform the trajectory of our project but also underscore the vast potential for future innovations in digital security.

## 3. Methodology

This project takes a bold step forward by cleverly combining the sophisticated field of steganography with well-established cryptographic techniques, all the while utilizing artificial intelligence advancements—more especially, the tactical use of Generative Adversarial Networks (GANs). This well-managed project progresses through multiple painstakingly planned phases, creating a strong basis for safe channels of communication.

Initially, the project thoroughly investigates important cryptographic techniques, setting the foundation for safe communication methods. One of these is the Key Exchange Protocol of Diffie-Hellman, that permits the private exchange of cryptographic keys in public spaces, enabling safe communication across open channels. The research also investigates the RSA Algorithm, which is a crucial component of public-key cryptography and guarantees that data traveling over a network may only be decoded by the intended recipient. It does this by converting data into a secure coded format using keys that are obtained from large prime integers. The project also explores the Elgamal Algorithm, which uses asymmetric key encryption to safeguard transmitted data using both public and private keys, improving security.

Advancing further, the initiative introduces innovative steganographic approaches. It looks at Least Significant Bit (LSB) Steganography, a brilliant technique that conceals large amounts of data by altering the smallest noticeable bits of pixel values in a picture. The changes are almost undetectable. Furthermore, the initiative explores GAN-based Steganography, utilizing Generative Adversarial Networks to embed hidden data within images. This approach not only secures the data but also ensures that the images used as carriers closely resemble authentic ones, effectively evading detection mechanisms.

A notable feature of this initiative is the seamless integration of cryptographic encryption with steganographic techniques, achieved through GAN architectures. This fusion not only encrypts the data but also conceals its presence, providing a dual layer of security. The encrypted information is so adeptly embedded within digital media that its existence remains known exclusively to the communicating entities.

Examining all steganographic Computational technique in use is vital for refining the initiative's methodology. Metrics like Mean Squared Error (MSE) and Peak Signal-to-Noise Ratio (PSNR) are used by the program to evaluate each method's effectiveness and determine which works best for different types of data and communication settings. This analysis procedure makes it possible to tailor security measures to specific communication demands, protecting the reliability and privacy of the sent data.

By amalgamating cryptographic methods, steganographic techniques, and AI innovations, the initiative develops a comprehensive, forward-looking framework for secure digital communications. This framework not only signifies the convergence of various technological fields but also establishes itself as a premier solution to the contemporary challenges of maintaining digital privacy and security in an interconnected world.

The following cryptographic algorithms considered are as follows: (i) Key Exchange Protocol of Diffie-Hellman (ii) RSA Algorithm, (iii) Elgamal Algorithm.

Along with all the cryptographic algorithms mentioned above, a comparative analysis of RSA and Elgamal Algorithm is performed.

## 3.1 Key Exchange Protocol of Diffie-Hellman

The efficacy of the Diffie-Hellman Computational technique relies on the complexity associated with solving discrete logarithm challenges, was introduced in the influential paper by Diffie and Hellman. This algorithm plays a vital role in securing communications across vulnerable networks, like the internet, by facilitating the secure exchange of encryption keys. The strength of Diffie-Hellman technique is based on resolving the hurdles behind the Discrete logarithm hurdle, using a prime number's primitive root to produce unique powers modulo p. This concept defines the discrete logarithm, crucial for cryptographic operations, expressed as *$dlog_{a,p}(b)$*. This procedure entails calculating a particular exponent *i* for which *$b = a^i \mod p$* is given an integer b and primitive root a of prime number *p*, where *$0 \leq i < (p-1)$*.

Here's how the Diffie–Hellman algorithm works:

Steps:

Adam and Barbie choose a prime number $p$ and a base $g$, where $g$ is a primitive root modulo $p$. These parameters are public and can be openly communicated.

First, let's explore the equations for crafting a confidential key and a shared key.

> $A(\text{Public key of Adam}) = g^a \bmod p$
> $B(\text{Public key of Barbie}) = g^b \bmod p$
> $S_A(\text{Shared Secret Key Generation of Adam}) = B^a \bmod p$
> $S_b(\text{Shared Secret Key Generation of Barbie}) = A^b \bmod p$
> Where $g$= base , $a$=Private key of Adam, $b$ = private key of Barbie

Public Key Generation:

Each party generates their public key:

Adam chooses a stochastic private key $a$ and computes $A$

Barbie chooses a stochastic private key $b$ and evaluates $B$.

Public Key Exchange:

Adam and Barbie exchange their public keys with each other:

Adam sends his public key $A$ to Barbie.

Barbie sends her public key $B$ to Adam.

Shared Secret Key Generation:

Adam calculates the shared secret key $S_A$.

Barbie calculates the shared secret key as $S_B$

Both Adam and Barbie now have the same shared secret key $S_{A/B}$, which they can use for symmetric encryption of their messages. The brilliance of the Diffie-Hellman method is that, while the public keys A and B are exchanged publicly, an eavesdropper cannot readily compute the shared secret key without knowing either a or b, which are kept private.

Essentially, the Diffie-Hellman algorithm enables two users to establish a common secret key through an unreliable channel , without the key being transmitted. The challenge of solving logarithmic calculations in a finite field makes it practically unattainable to figure out the shared secret key without the private keys, securing the exchange process.

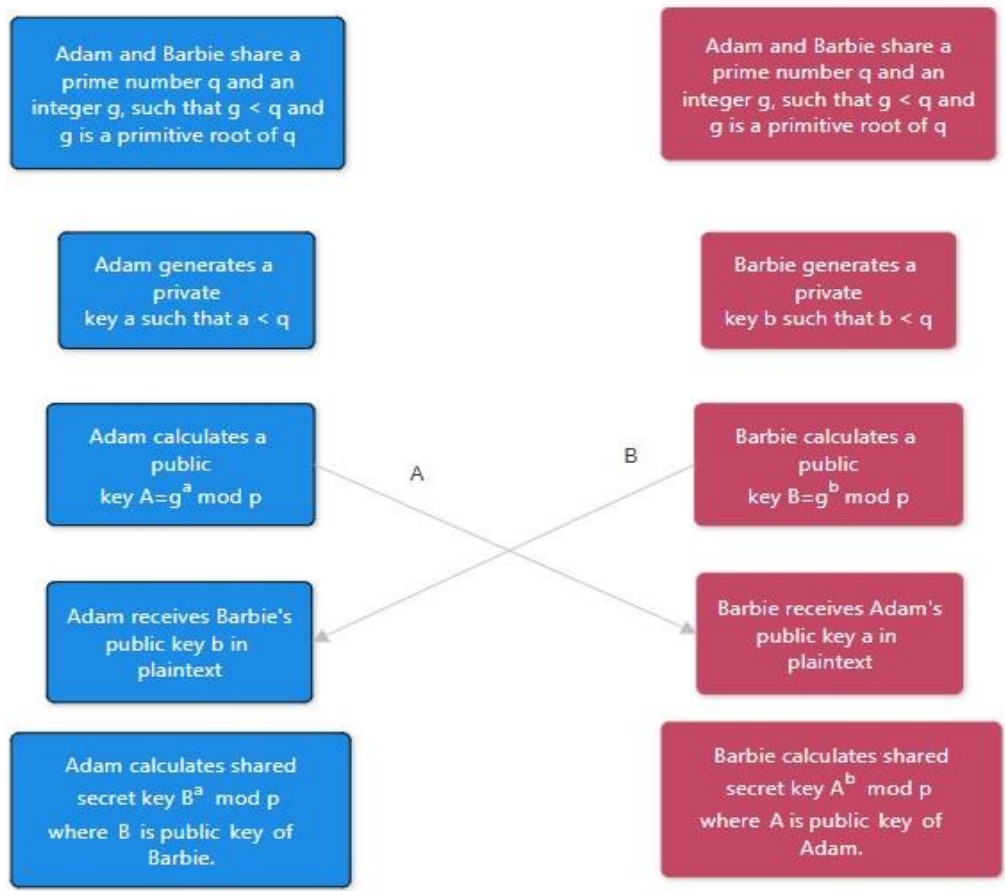

Fig 1: Illustration of the working principle of Diffie Hellman Key Exchange

## 3.2 RSA Algorithm

The RSA algorithm utilizes an encryption system where plaintext and ciphertext are represented as integers within the range of *0 to (n-1)*. Typically, *(n)* is 1024 bits in size, or about 309 decimal digits.

RSA uses power-based calculations to encrypt data in blocks, with each less than a number *(n)*, making the maximum block size ($log_2 n + 1$) bits; commonly, blocks are *(i)* bits.

where *2i < n <= 2i+1*.
The RSA algorithm processes encryption and decryption via specific mathematical operations:

- *(C)* is the encrypted result, calculated as *(C = Me mod n)*.
*M = Cd mod n = (Me)d mod n = Med mod n*
- *(M)* is the original message, obtained by decrypting *(C)* as *( M = Cd mod n)*, which mathematically equals *(M = (Me)d mod n )*.

Both the sending and receiving parties need to know *(n)*. The sender has the encryption key *(e),* while only the recipient knows the decryption key *(d)*. This framework defines RSA as an asymmetric encryption method, with the public key being *PU = {e, n}* and the private key as *PR = {d, n}*. For effective use in asymmetric encryption, RSA must satisfy these criteria:

- There must be a way to choose e, d, and n so that *(Med mod n)* returns M for all *(M < n)*.
- Calculating *(C = Me mod n)* and *(Cd mod n)* should be straightforward for any *(M < n)*.
- It should be computationally impractical to deduce (d) from (e) and (n).

The relationship *(Med mod n = M)* stands if e and d are inverses modulo the totient function of Euler's of n, denoted as *F(n)*.
For primes *p* and *q*,
*F(pq) = (p - 1)(q - 1)*
and the connection between e and d is defined by the equation.

*ed mod F(n) = 1.*

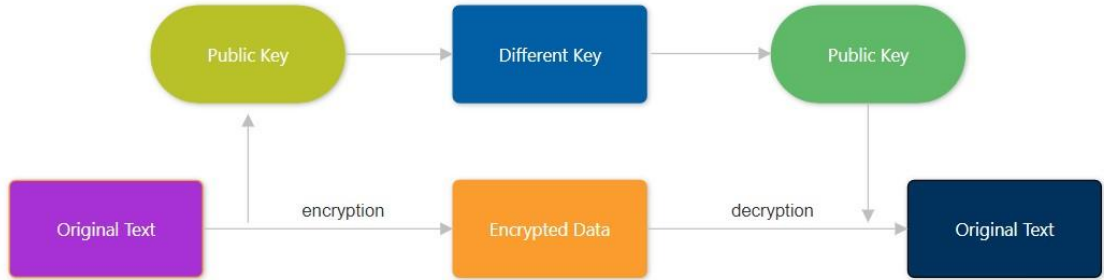

Fig 2: The RSA Algorithm

Key Generation Example:

- Select two prime numbers. For this instance, let's take *p = 13 and q = 19*.
- Calculate the prime's product, *(n = p \* q) = (13\*19) = 247*.
- Determine totient function of Euler's for *n, F(n) = (p-1)\*(q-1) = (12\*18) = 216)*.
- Choose e so that it is a coprime of *F(n)* and less than *F(n)*. Here, we'll select *(e = 11)*.
- Find d so that d times e is congruent to *1 modulo F(n)* and d is less than *F(n)*. An appropriate value for *d* is *(d = 59) as ( 59 times 11 mod 216 = 1)*.
- This process yields the public key *( PU = {11, 247})* and the private key *( PR = {59, 247})*.

Ciphering and Deciphering Illustration:
For a plaintext: *(M = 65)*
Ciphering: Compute *(C = $M^e$ mod n = $65^{11}$ mod 247)* to get the ciphertext *(C)*.
Deciphering: Compute *(M = $C^d$ mod n = $C^{59}$ mod 247)* to retrieve the plaintext *(M = 65)*.

## 3.3 Elgamal Algorithm

T. Elgamal created a public-key encryption algorithm based on Diffie-Hellman principles in 1984, which has since been included into various standards, including the Digital Signature Standard (DSS) and the Secure/Multipurpose Internet Mail Extensions (S/MIME) email standard. The Elgamal encryption scheme relies on a prime number *(q)* and a primitive root modulo *(q)*. Here's how user A would generate a key pair, in addition to how the Elgamal system handles encryption and decryption:

Key Establishment by User A:

- Choose a secret integer *($S_A$)* where *($1 < S_A < r - 1$)*, with 'r' being a prime.

- Compute *($K_A = g^{S_A} \bmod r$)*. $K_A$ denotes the computed public value that User A derives from their secret key $S_A$.

- Keep *($S_A$)* as A's secret key and publicize *{r, g, $Z_A$}* as A's public credentials.

Message Encryption by Any User B:

- Convert the plaintext into an integer *(P)* between *0 and (r - 1)* If the text is lengthy, segment it into blocks, each block being an integer smaller than *(r)*.

- Select a confidential integer *(l)* within the range of *1 to (r - 1)*.

- Create a session key *($Sk = Z_A^l \bmod r$)*.

- Formulate the encrypted message as a numeric pair *(E1, E2)* where:

  *($E1 = g^l \bmod r$)*

  *($E2 = Sk * P \bmod r$)*.

Message Decryption by User A:

- Derive the session key *($Sk = E_1^{S_A} \bmod r$)*.

- Retrieve the plaintext *P* by computing $P = (E2 * E_1^{-1}) \bmod r$.

Here, *(SA)* is User A's private key, while *(ZA)* is part of User A's public key set along with *(r)* and *(g)*. User B utilizes *(Sk)* for encrypting the plaintext *(P),* resulting in the ciphertext components *(E1)* and *(E2)*.

This process demonstrates the effectiveness of the Elgamal scheme.

## 3.4 Comparative Study of RSA and Elgamal Algorithm

Table 1: Comparison of RSA and Elgamal Algorithm (Arhin Jnr, 2023)

| | RSA | | Elgamal | |
|---|---|---|---|---|
| **Key Size in Bits** | **Encryption** | **Decryption** | **Encryption** | **Decryption** |
| 1024 | 0.139 | 0.231 | 0.703 | 0.305 |
| 2048 | 0.561 | 0.631 | 1.107 | 0.705 |

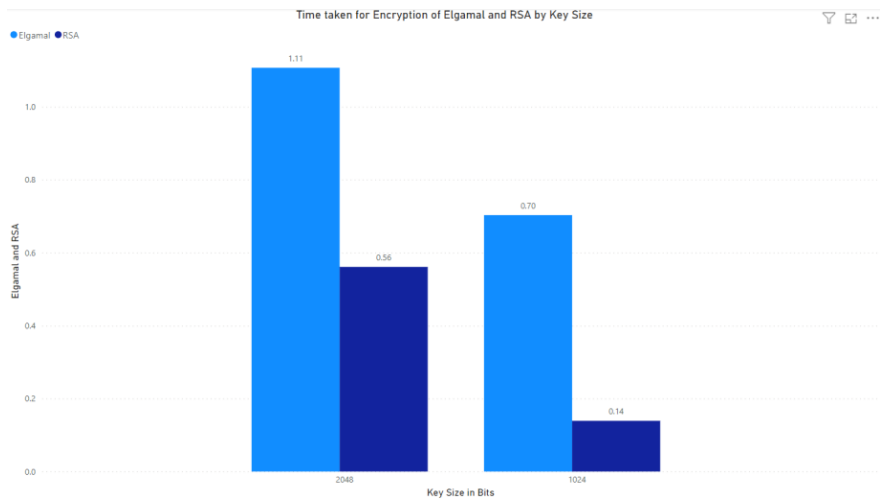

Fig 3: Displaying the Encryption speed difference using key size of 1024 bits.

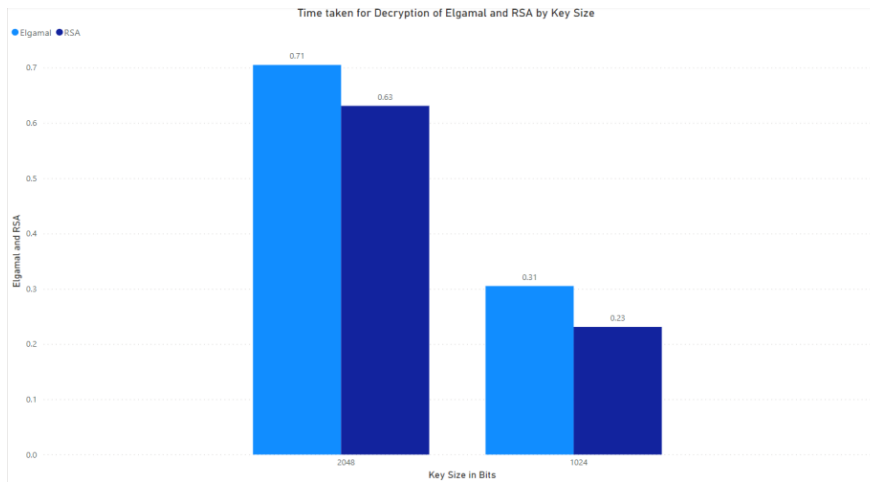

Fig 4: Displaying the encryption and decryption speed difference using key size of 2048 bits

In an analysis of the RSA and Elgamal cryptographic algorithms implemented in Python, it was found that RSA consistently encrypts and decrypts text faster than Elgamal. This difference in speed is attributed to Elgamal's encryption and decryption requiring multiple modular exponentiations, in contrast to RSA's single exponentiation operation. Although Elgamal's decryption is faster than its encryption due to the latter's more complex calculations involving random number generation and additional modular exponentiations, RSA remains superior in both speed aspects.

The fight against unlawful online digital material copying, hacking, illegal interception, and confidentiality violations has gotten more intense. Technologies such as cryptography, steganography, watermarking, and fingerprinting are deployed to safeguard communication secrecy. Cryptography involves encrypting data to obscure it from unauthorized access, while steganography conceals secret information within a cover object, making it undetectable to intruders. Unlike encryption, which only obscures information meaning, steganography ensures hidden information, enhancing security measures

When assessing the effectiveness of steganographic techniques, various metrics come into play for comparison. Mean Squared Error (MSE) and Peak Signal-to-Noise Ratio (PSNR) are two metrics that are frequently used. Higher PSNR values indicate better image quality preservation. PSNR measures the ratio between the maximum potential signal power and the power of noise impacting its fidelity.

MSE, conversely, computes the average squared difference between the stego component and the starting wrap component, offering a quantitative assessment of distortion introduced during embedding. Lower MSE values signify reduced distortion and more effective concealment of hidden information within the cover object.

By employing metrics such as PSNR and MSE, researchers can gauge and compare the performance of various steganographic techniques, ultimately facilitating the emergence of more trustworthy and safeguarded techniques to hide delicate data in electronic mediums.

The mean squared error (MSE) is the most straightforward method to define PSNR. MSE is defined as follows: given a stochastic-free m×n monotonous picture I and its stochastic estimation K,

The following steganographic algorithms considered are as follows: (i) LSB (Least Significant Bit) based Text Steganography (ii) GAN based steganography, (iii) Elgamal Algorithm.

The MSE is defined as

$$\text{MSE} = \frac{1}{mn}\sum_{i=0}^{m-1}\sum_{j=0}^{n-1}[I(i,j) - K(i,j)]^2$$

The PSNR (in dB) is defined as

$$\begin{aligned} \text{PSNR} &= 10 \log_{10} \frac{MAX_I^2}{MSE} \\ &= 20 \log_{10} \left(\frac{MAX_I}{\sqrt{(MSE)}}\right) \\ &= 20 \log_{10} (MAX_I) - 10 \log_{10} (MSE) \end{aligned}$$

The maximum pixel value of the image is denoted by $MAX_I$ in this case. This equals 255

when the pixels are represented with 8 bits per sample. In general, $MAX_I$ equals $2B - 1$ when samples are encrypted using linear PCM with B bits per sample.

## 3.5 LSB (Least Significant Bit) based text steganography

Our primary research focus lies within the spatial domain, which involves manipulating the original image's component regions to incorporate further data.

Within this domain, we specifically delve into the LSB (Least Significant Bit) technique. LSB entails embedding each bit of data, be it characters or images, into the least significant portion of the imagery on the cover.This ensures that the alterations introduced during the insertion process are imperceptible to the human eye. Our investigation predominantly revolves around the LSB technique, wherein we conceal varying lengths of secret data within cover images. Next, we do an assessment by measuring the difference in the preliminary and secured photos' Mean Square Error (MSE) and Peak Signal to Noise Ratio (PSNR).

LSB Encoder

- Start with the introductory picture.
- Transform the image to RGBA mode, which uses Red, Green, Blue, and Alpha (transparency) channels for each of them.
- Convert the text message to binary format, using 8 bits for each character.
- The binary message should be encapsulated into the least significant bit (LSB) of every pixel in the image's Red, Green, and Blue color regions.
- Save the modified image with the embedded message.
- The resulting image now contains the hidden message using LSB steganography with 8 bits per character for encoding.

LSB Decoder

- Start with the image containing the hidden message.
- Extract the LSB of each color channel of each pixel to reconstruct the binary message.
- Convert the binary message back to text format, character by character, using 8 bits per character.
- Stop decoding when the delimiter ("#####") marking the end of the message is encountered.

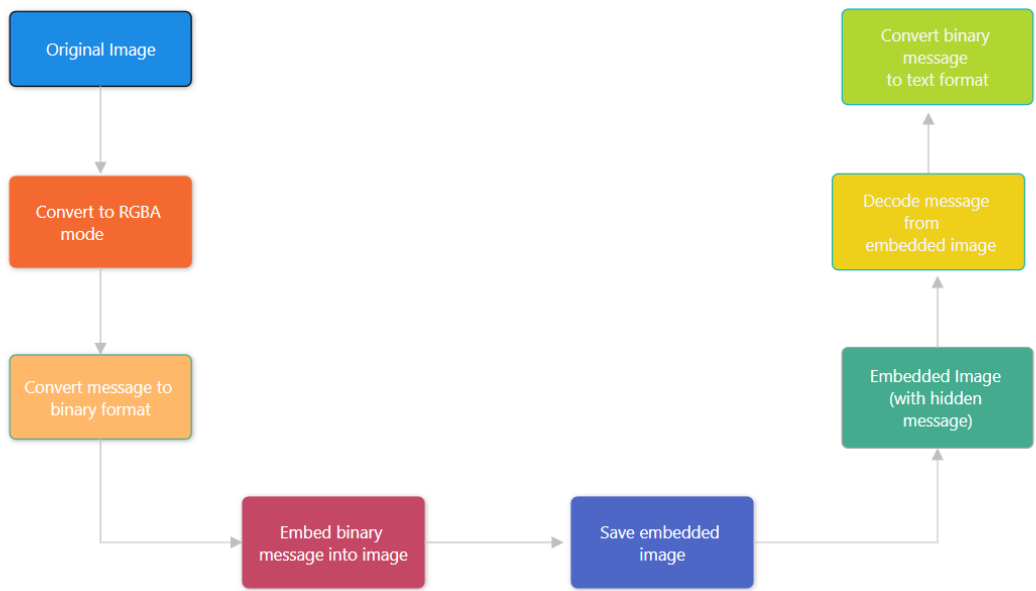

Fig 5: LSB Based Text Steganography Architecture

Experimental Results:

Table 2: PSNR(dB) and MSE values for LSB Method

| **Image(799x792).jpg** | **PSNR RATIO (dB)** | **MSE** |
|---|---|---|
| **Barbara** | 93.839034 | 2.69E-05 |
| **Cat** | 95.729596 | 1.74E-05 |
| **Cameraman** | 93.355987 | 3.00E-05 |
| **Boat** | 95.232719 | 1.95E-05 |

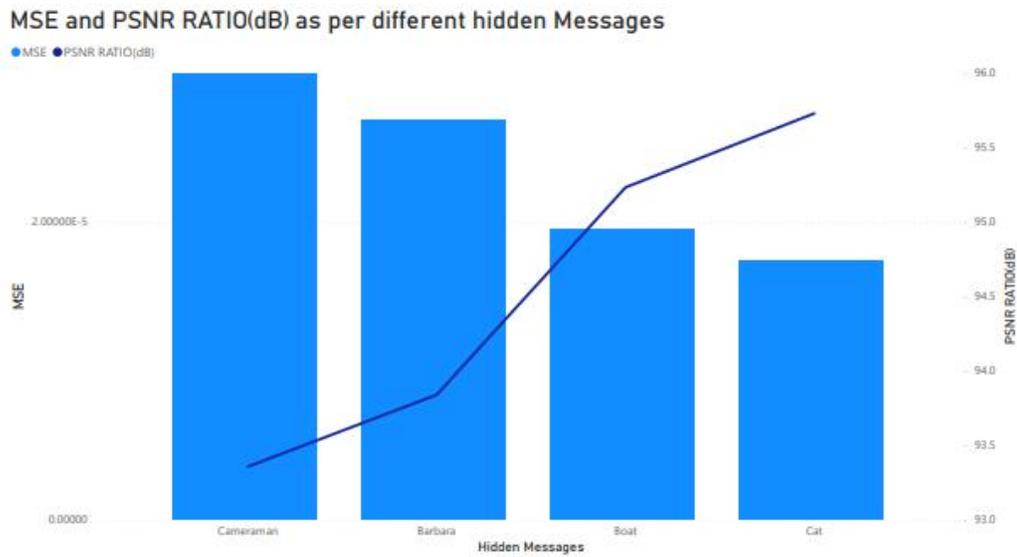

Fig 6: MSE and PSNR RATIO (dB) as per different messages

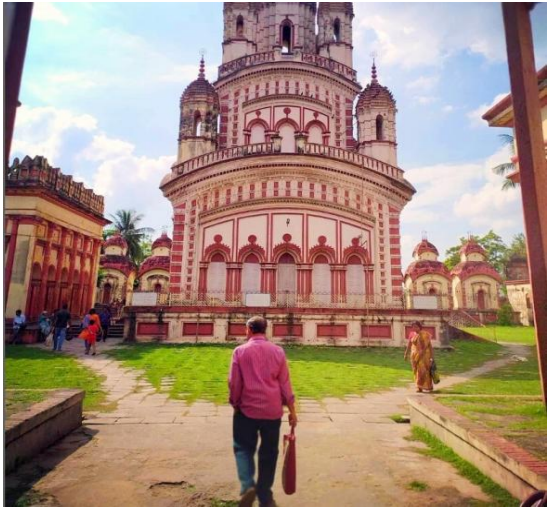 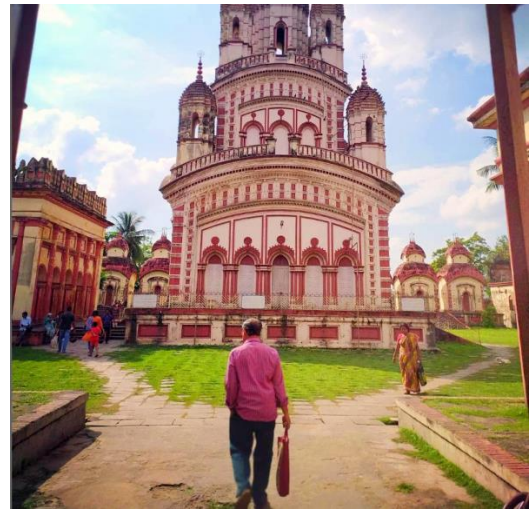

(a) Before embedding hidden message.          (b) After embedding hidden message.

Fig 7: An image before and after Steganography

LSB steganography is implemented such that a message is embedded into an image and then decode it. It converts the image to RGBA format, converts the message to binary, embeds the binary message into the inserts a delimiter that denotes the end of the message in the least significant bit of each color channel in the image, saves the modified image, decodes the message from the modified image using the LSB of each color channel, and calculates the PSNR and MSE.

Using the Mean Square Error (MSE) and Peak Signal to Noise Ratio (PSNR), we analyzed an identical image with various text lengths. A larger PSNR value denotes an enhanced image.

Results:

The *Cat* image has the highest PSNR (95.729596 dB) and the lowest MSE (1.74E-05), which suggests that the steganographic process produced the least distortion in this image, making it the best carrier among the four for embedding hidden messages.

The *Cameraman* image has the lowest PSNR (93.355987 dB) and the highest MSE (3.00E-05), indicating that it experienced the most distortion during the process, making it the least suitable carrier for hiding information without detection.

The *Boat* image, with a PSNR of 95.232719 dB and an MSE of 1.95E-05, and the *Barbara* image, with a PSNR of 93.839034 dB and an MSE of 2.69E-05, both show intermediate performance in terms of quality and error, with the *Boat* image leaning towards better performance than *Barbara*.

## 3.6 GAN based steganography

CGANs are generative models that learn to create data samples under specific conditions, such as embedding secret information into images while preserving visual quality, known as CGAN-based steganography.

This technique involves training a CGAN with cover images and their corresponding secret messages. The generator produces images with hidden messages, while the discriminator distinguishes between these and real cover images, improving the generator's ability to hide information.

After training, the CGAN can embed secret messages into cover images by generating images with the hidden message, which recipients can extract using the trained CGAN. CGAN-based steganography balances security and visual quality, with effectiveness depending on factors like message capacity, detection robustness, and image quality.

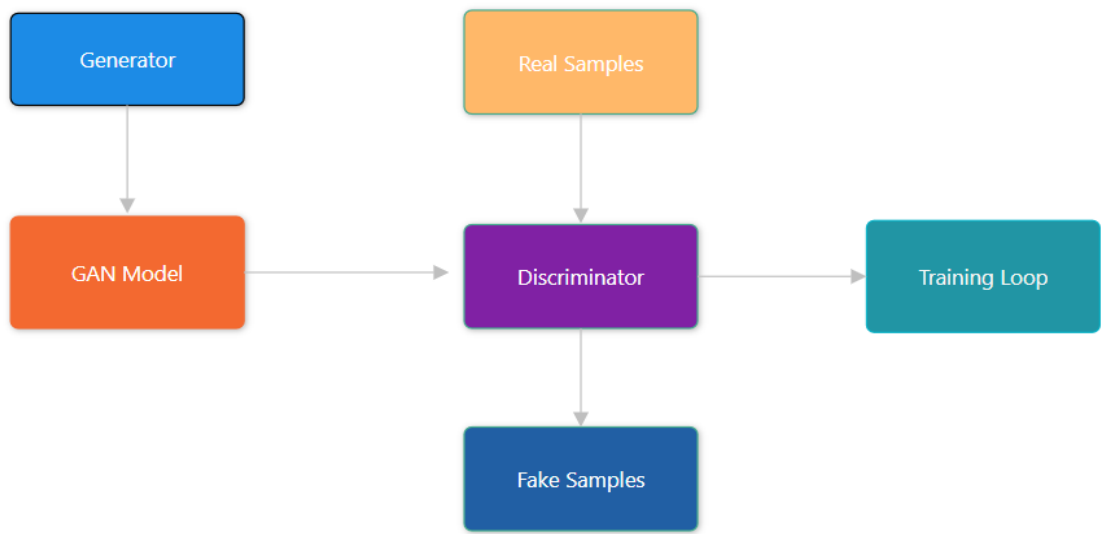

Fig 8: CGAN Training

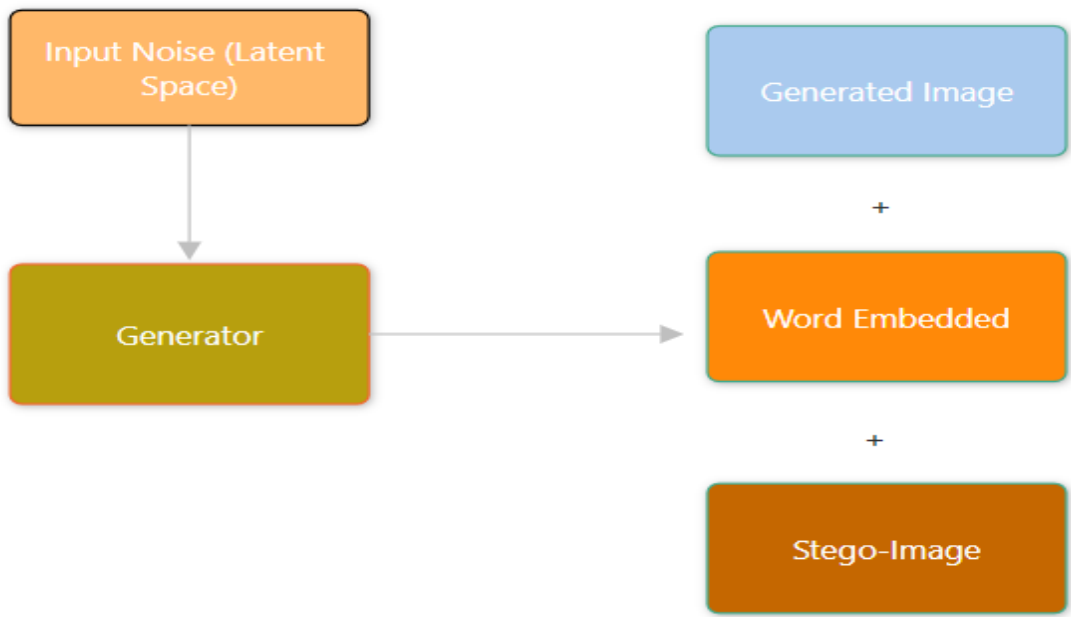

Fig 9: Steganography Implemented using the trained CGAN.

CGAN Training Process:

Generator:
Takes input noise (latent space).
Produces fake/generated images resembling real data.

Discriminator:
Takes both real samples (from the dataset) and fake samples (generated by the Generator).Discriminates between real and fake images, outputting the probability of an image being real.

GAN Model:
Combines the Generator and Discriminator in a training loop.
The Generator tries to create images that the Discriminator will classify as real.

The Discriminator learns to better distinguish between real and generated images. The training loop involves backpropagation and gradient descent to update the weights of both models.

Steganography Implementation:

Input Noise (Latent Space):

Random noise that serves as the seed for image generation in the Generator.

Generated Image:

The Generator uses the input noise to create a new image that tries to mimic the distribution of the real samples.

Word Embedding:

A word or message is embedded into the generated image using a steganography technique. The embedding is done in such a way that it is imperceptible to the human eye, preserving the visual quality of the image.

Stego Image:

The final output image that contains the embedded word/message. This image looks similar to the generated image but contains the hidden information.

Combined Workflow:

The CGAN is first trained with real and fake samples until it reaches a satisfactory level of performance. Once the CGAN is trained, it can be used to generate images into which words or messages can be embedded using steganography. The steganographic implementation utilizes the trained Generator to produce images that serve as carriers for the embedded information. The end product is a stego image, which appears like any other generated image but contains hidden data that can be extracted later with the proper decoding technique.

Experimental Results:

Table 3: PSNR(dB) and MSE values for CGAN based Method.

| Text | MSE | PSNR RATIO (dB) |
|---|---|---|
| **Barbara** | 0.001 | 78.130804 |
| **Cat** | 0.000333 | 82.902016 |
| **Cameraman** | 0.0013 | 77.99137 |
| **Boat** | 0.0005333 | 80.860816 |

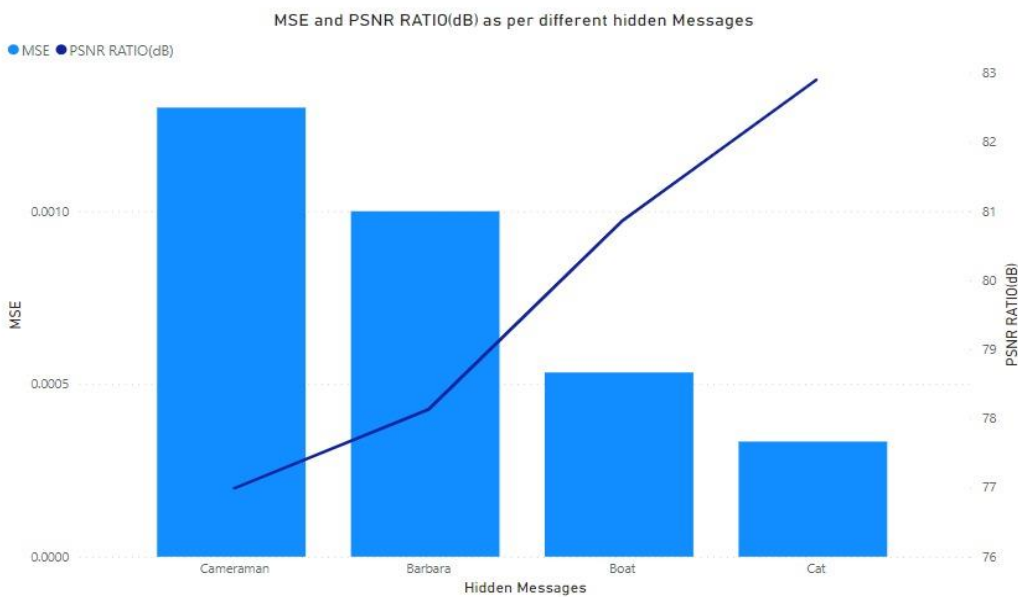

Fig 10: MSE and PSNR RATIO(dB) as per different messages

Results:

Barbara: Shows an MSE of 0.001 and a PSNR of 78.13084 db. This suggests a moderate level of error with a fairly high signal-to-noise ratio, indicating good quality of the hidden message's reconstruction.

Cat: Exhibits the lowest MSE (0.000333) and the highest PSNR (82.902016 dB), indicating the embedding process caused the least distortion and resulted in the highest quality reconstruction among the four.

Cameraman: Has a higher MSE (0.0013) and a lower PSNR (77.99137 dB) than Barbara, suggesting more error and a lower reconstruction quality.

Boat: Displays an MSE of 0.0005333 and a PSNR of 80.860816 dB, suggesting it has better performance than Barbara and Cameraman, but not as good as Cat.

The effectiveness of steganographic methods is influenced by the choice of the underlying images, as demonstrated by the differing MSE and PSNR readings. In this particular set of experiments, the 'Cat' image proved to be the most suitable medium for message concealment, evidenced by its minimal MSE and maximal PSNR, signifying superior preservation and extraction of the embedded data. On the other hand, the 'Cameraman' image resulted in the most significant distortion and the poorest message retrieval quality, as shown by its highest MSE and lowest PSNR.

These findings highlight the importance of careful image selection in steganography, given its substantial impact on the concealment and integrity of encoded messages. For steganography to be effective, it's crucial to utilize images that lead to a low MSE and a high PSNR, ensuring the embedded data remains undetectable and intact when extracted. Selecting images that achieve these criteria is vital for the success of steganographic techniques.

## 3.7 Hybrid Secure Messaging: Cryptography Meets Steganography

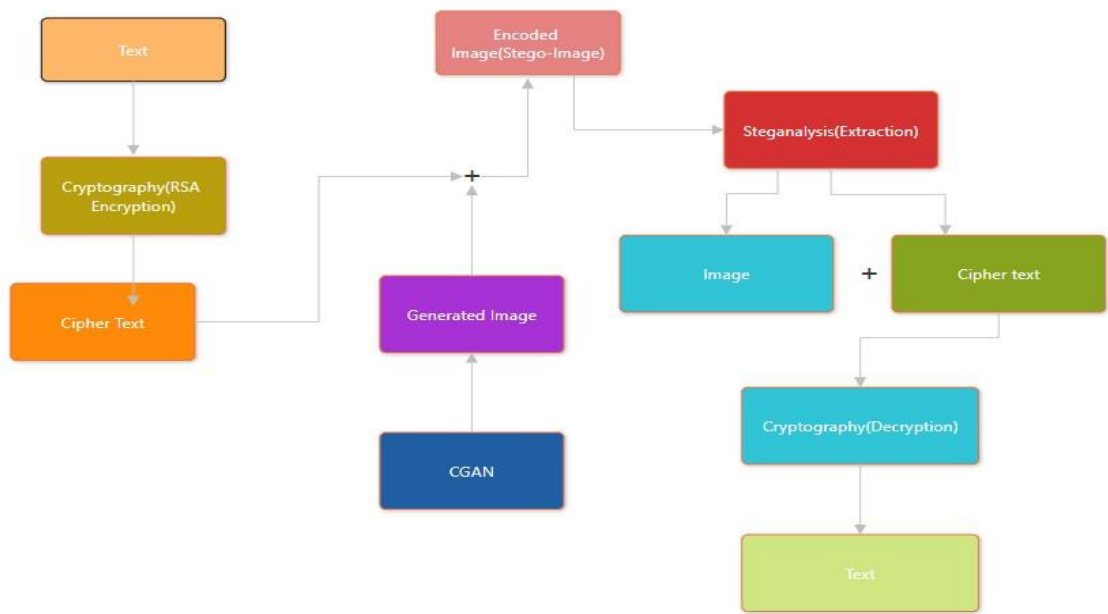

Fig 11: Concatenated GAN Architecture of Cryptography and Steganography

Input Image:

The initiation of the process involves an original image that will serve as the carrier for the concealed message.

Steganography (Embedding):

During this phase, a text message, encoded in an 8-bit format—meaning each character is represented by 8 bits or one byte—is intricately woven into the input image. This is done with precision to ensure the visual integrity of the image is maintained, rendering the modifications virtually imperceptible.

Cryptography (Encryption):

The text message undergoes encryption using the RSA algorithm at a point before or following the steganographic embedding. RSA employs a public key for the encryption of data, rendering it decryptable solely with the matching private key. Although this step does not visibly alter the image, it effectively secures the embedded message, guaranteeing its confidentiality should the steganography be uncovered.

Encoded Image:

The outcome of this process is an image that, to the casual observer, appears unchanged from the original but now carries within it the covertly embedded and encrypted message.

cGAN Implementation:

To create an image that retains the message's encrypted nature while mimicking the original's aesthetic traits, a Conditional Generative Adversarial Network (cGAN) is adopted. The cGAN is specifically trained to generate outcomes based on defined input criteria—in this scenario, to generate an image that effectively hides the message while remaining visually faithful to the original.

Steganography (Extraction):

The recipient of the image will then engage in extracting the 8-bit encoded data from the image, a process that necessitates knowledge of the particular steganography technique utilized for embedding.

Cryptography (Decryption):

Post-extraction, the still-encrypted message must be decrypted using the RSA private key. This step is crucial for secure communication, as it ensures that only the communication can be intercepted and deciphered by the possessor of the private key.

Image and Text Decoded:

In the concluding step, the intended recipient is presented with both the original image and the now decrypted text message. The phrase "Decoded Image" at this stage could be more aptly termed as "Decrypted Message," since the primary aim is the revelation of the concealed text.

Experimental Results:

Table 4: PSNR(dB) and MSE values for concatenated architecture

| Text | MSE | PSNR RATIO (dB) |
|---|---|---|
| **Barbara** | 0.033767 | 62.84592 |
| **Cat** | 0.0351 | 62.67773 |
| **Cameraman** | 0.0355 | 62.62852 |
| **Boat** | 0.034433 | 62.76101 |

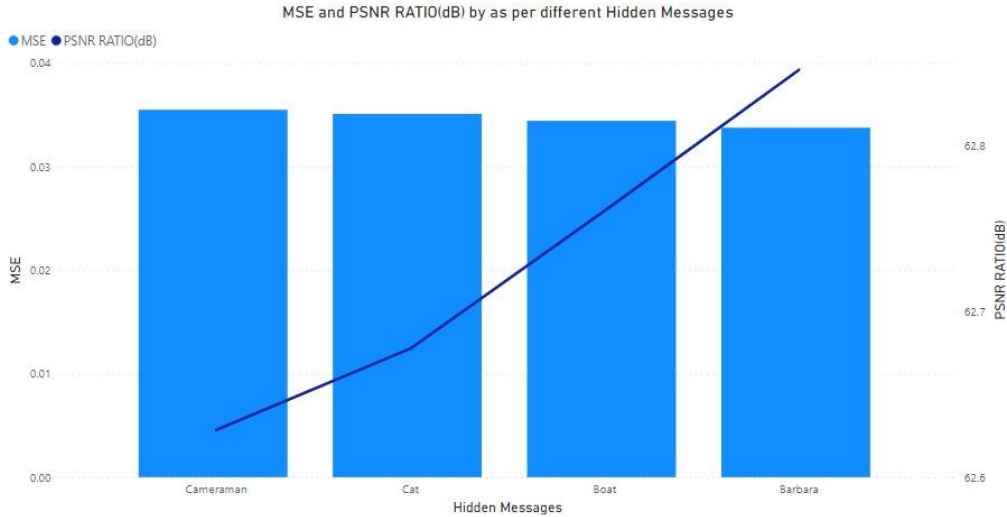

Fig 12: MSE and PSNR RATIO(dB) as per different messages

All the images exhibit high PSNR values, which implies that the reconstruction or compression process preserves the image quality well.

considering the high PSNR values and low MSE values, it can be concluded that the quality of the projected images is remarkably comparable to that of the genuine photos.The slight negative correlation between MSE and PSNR observed in the graph is consistent with their definitions: as error increases, quality typically decreases.

Since the differences in MSE and PSNR are small, it suggests that the process applied to these images is quite robust, yielding consistently good results across different types of images.

## 4. Results And Analysis

Table 5: PSNR(dB) and MSE values for all the steganographic methods

| Image(799x792).jpg | LSB | | CGAN | | Concatenated | |
|---|---|---|---|---|---|---|
| | PSNR RATIO (dB) | MSE | PSNR RATIO (dB) | MSE | PSNR RATIO (dB) | MSE |
| Barbara | 93.839034 | 0.0000269 | 78.130804 | 0.001 | 62.84592 | 0.033767 |
| Cat | 95.729596 | 0.0000174 | 82.902016 | 0.000333 | 62.67773 | 0.0351 |
| Cameraman | 93.355987 | 0.00003 | 77.99137 | 0.0013 | 62.62852 | 0.0355 |

The table provides PSNR and MSE values for three different steganographic algorithms applied to three images (Barbara, Cat, and Cameraman). Additionally, the Concatenated method is noted to include CGAN with RSA encryption.

Analyzing the numerical values in (i) LSB (Least Significant Bit) Based Steganography (ii) cGAN Based Steganography, (iii) Hybrid architecture Based Steganography.

## 4.1 LSB (Least Significant Bit) Based Steganography

- Yields the highest PSNR values (93.83 for Barbara, 95.72 for Cat, 93.35 for Cameraman), indicating the stego-images retain high similarity to the original images, suggesting minimal quality degradation.
- Has the lowest MSE values, indicating minimal error between the stego-images and the original images.

Use Cases for LSB:
- Situations requiring high-quality stego-images where minimal distortion is paramount.Scenarios where the hidden data volume is relatively low, as LSB can be more susceptible to image manipulation or compression.

## 4.2 cGAN Based Steganography
- Shows lower PSNR values than LSB, indicating more noticeable image degradation after data embedding.
- MSE values are higher than LSB but still relatively low, suggesting a moderate level of error.

Use Cases for CGAN:
- Environments where some degradation of image quality is acceptable for enhanced security.
- Applications that may benefit from the adversarial nature of GANs, which could potentially make the steganography more robust against detection methods.

## 4.3 Hybrid architecture Based Steganography

- Has significantly lower PSNR values, suggesting that when juxtaposed to the original, the stego-image's quality has significantly deteriorated.
- Exhibits the highest MSE values among the three methods, indicating the highest level of error due to data embedding.

Use Cases for Concatenated Method:

- Use in high-security applications where the encryption of the payload is crucial, and image quality is a secondary concern.
- Situations where both the robustness against steganalysis and the confidentiality of the embedded data are of utmost importance, such as in sensitive communications.

In conclusion:

- The LSB algorithm appears best suited for applications where image quality is a priority, and the risk of steganalysis is minimal.
- The CGAN algorithm might be preferable when there is a need for a balance between image quality and security against detection.
- The Concatenated method seems tailored for scenarios that demand high security for the embedded data at the cost of image quality, which could be compensated for by the robustness provided by the incorporation of CGAN with RSA encryption.

## 5. Conclusion and Future Work

The conclusion of the paper you provided highlights that while Generative Adversarial Networks (GANs) can produce an unlimited number of cover images, there are challenges in ensuring these images are natural enough to effectively conceal secret messages. The study found that when the generator within a GAN is well-trained, the discriminator, intended to filter out unnatural images, becomes relatively weak. This observation suggests that for the images generated by the GAN to be effectively used as steganography cover images, the discriminator needs to be improved through the application of a new loss function. This need for an enhanced discriminator presents a critical area for future research.

Future work will focus on developing and testing new loss functions for the discriminator, aiming to improve its ability to evaluate the naturalness of images produced by the generator. This advancement is vital for the proposed steganography framework, which relies on GANs to produce cover images that are indistinguishable from natural images, thereby enhancing the security of hidden messages.

The paper discusses various challenges conceivable prospects for imagery steganography investigation in the coming years, particularly in leveraging deep learning technologies. These potential future directions include:

Exploring Other Network Architectures: While Generative Adversarial Networks (GANs) and Convolutional Neural Networks (CNNs) are widely used, there's potential in exploring other architectures like Recurrent Neural Networks (RNNs) for steganography. Customizing and experimenting with different network architectures, including variations of GANs, could lead to more efficient and secure steganography methods.

2. Expanding Secret Information Types: Most current methods focus on hiding text or grayscale images. There's a need for research into steganography techniques that can hide different types of media, including color images and videos, within cover media without significant loss of quality or detectability.

3. Optimization and Efficiency: Investigating ways to optimize deep learning models for steganography to reduce training times, improve computational efficiency, and decrease storage requirements. This could make steganography more practical for real-world applications where resources are limited.

4. Quantum Computing: As quantum computing continues to evolve, exploring its application in steganography could offer new methodologies for securing data. Quantum steganography could potentially offer unprecedented security levels due to quantum computing's inherent properties.

5. Hybrid Approaches: Combining traditional steganography methods with deep learning could offer a balance between the security of traditional methods and the robustness of deep learning models. Hybrid approaches may also mitigate some deep learning models' weaknesses in steganography.

6. Benchmark Datasets and Evaluation Metrics: Establishing benchmark datasets specific to steganography could standardize evaluation and comparison across different methods. Similarly, developing comprehensive evaluation metrics that consider security, capacity, and robustness could provide a clearer picture of a method's effectiveness.

7. Security Against Attacks : Future research could also focus on enhancing the security of steganography methods against various attacks, including man-in-the-middle attacks and image tampering. Developing methods that can withstand these attacks is crucial for the practical application of steganography in secure communications.

By addressing these challenges and exploring these future directions, the field of image steganography can continue to advance, offering more secure and efficient methods for hiding information within images .